\documentclass[12pt,a4paper]{article}
\begin{document}
\textwidth=135mm
 \textheight=200mm
\begin{center}
{\bfseries The property of maximal transcendentality
in the  ${\mathcal N}=4$ SYM
\footnote{{\small Talk at
International Bogolyubov Conference
``Problems of Theoretical and Mathematical Physics", JINR, Dubna,
August 21 - 27, 2009.}}}
\vskip 5mm
A.V. Kotikov$^{\dag}$
\vskip 5mm
{\small {\it $^\dag$ BLTP, Joint Institute for
Nuclear Research, 141980 Dubna, Russia}}
\\
\end{center}
\vskip 5mm
\centerline{\bf Abstract}
We show
results for the universal anomalous dimension
$\gamma _{uni}(j)\,$ of Wilson twist-2 operators in the
${\mathcal N}=4$ Supersymmetric Yang-Mills theory in the first
three orders of perturbation theory.
These expressions are obtained by extracting the most complicated contributions
from the corresponding anomalous dimensions in QCD.
\vskip 10mm
\section{\label{sec:intro}Introduction}
The anomalous dimensions (ADs) of the twist-2 Wilson operators govern the
Bjorken scaling violation for parton distributions in a framework of Quantum
Chromodynamics (QCD) \cite{DGLAP,Lund}.
Now they
are known up to the
next-to-next-to-leading order (NNLO) of the perturbation theory~\cite{VMV}.

The QCD expressions for ADs
can be transformed to the case
of the
${\mathcal N}$-extended Supersymmetric Yang-Mills theories (SYM) ~\cite{BSSGSO}
if one will use for the Casimir operators $C_{A},C_{F},T_{f}$ the following
values $C_{A}=C_{F}=N_{c}$, $T_{f}n_f={\mathcal N}N_{c}/2$.
For ${\mathcal N}\!\!=\!\!2$ and ${\mathcal N}\!\!=\!\!4$-extended SYM the
ADs of the Wilson operators get also
additional contributions coming from scalar particles~\cite{KL}.
These ADs
were calculated in the next-to-leading order
(NLO)~\cite{KL,KoLiVe}
for the ${\mathcal N}=4$ SYM.

However, it turns out, that the expressions
for eigenvalues of the AD
matrix in the ${\mathcal N}=4$ SYM can be derived directly
from the QCD anomalous dimensions without tedious calculations by using a
number of plausible arguments. The method elaborated in Ref.~\cite{KL} for
this purpose is based on special properties of
the integral kernel for
the Balitsky-Fadin-Kuraev-Lipatov (BFKL) equation~\cite{BFKL}-\cite{KL00}
 in this model
and a new relation between the BFKL and
Dokshitzer-Gribov-Lipatov-Altarelli-Parisi (DGLAP) equations (see~\cite{KL}).

\section{Leading order AD
matrix in ${\mathcal N}=4$ SYM}

In the ${\mathcal N}=4$ SYM theory~\cite{BSSGSO}
one can introduce the following colour and $SU(4)$ singlet local Wilson twist-2
operators~\cite{KL,KoLiVe}:
\begin{eqnarray}
\mathcal{O}_{\mu _{1},...,\mu _{j}}^{g} &=&\hat{S}
G_{\rho \mu_{1}}^{a}{\mathcal D}_{\mu _{2}}
{\mathcal D}_{\mu _{3}}...{\mathcal D}_{\mu _{j-1}}G_{\rho \mu _{j}}^a\,,
\label{ggs}\\
{\tilde{\mathcal{O}}}_{\mu _{1},...,\mu _{j}}^{g} &=&\hat{S}
G_{\rho \mu_{1}}^a {\mathcal D}_{\mu _{2}}
{\mathcal D}_{\mu _{3}}...{\mathcal D}_{\mu _{j-1}}{\tilde{G}}_{\rho \mu _{j}}^a\,,
\label{ggp}\\
\mathcal{O}_{\mu _{1},...,\mu _{j}}^{\lambda } &=&\hat{S}
\bar{\lambda}_{i}^{a}\gamma _{\mu _{1}}
{\mathcal D}_{\mu _{2}}...{\mathcal D}_{\mu _{j}}\lambda ^{a\;i}\,, \label{qqs}\\
{\tilde{\mathcal{O}}}_{\mu _{1},...,\mu _{j}}^{\lambda } &=&\hat{S}
\bar{\lambda}_{i}^{a}\gamma _{5}\gamma _{\mu _{1}}{\mathcal D}_{\mu _{2}}...
{\mathcal D}_{\mu_{j}}\lambda ^{a\;i}\,, \label{qqp}\\
\mathcal{O}_{\mu _{1},...,\mu _{j}}^{\phi } &=&\hat{S}
\bar{\phi}_{r}^{a}{\mathcal D}_{\mu _{1}}
{\mathcal D}_{\mu _{2}}...{\mathcal D}_{\mu _{j}}\phi _{r}^{a}\,,\label{phphs}
\end{eqnarray}
where ${\mathcal D}_{\mu }$ are covariant derivatives.
The spinors $\lambda _{i}$ and
field tensor $G_{\rho \mu }$ describe gluinos and gluons, respectively, and
$\phi _{r}$ are the complex scalar fields.
For all operators in Eqs.~(\ref{ggs})-(\ref{phphs}) the symmetrization of the tensors
in the Lorentz indices
$\mu_{1},...,\mu _{j}$ and a subtraction of their traces is assumed.
Due to the fact that all twist-2 operators belong to the same supermultiplet
the eigenvalues of AD
matrix can be expressed through
one {\it universal AD}
$\gamma_{uni}(j)$ with shifted
argument.
%
At the leading order (LO), it has the
form (\ref{uni1.1}) \cite{LN4}.

\section{Transcendentality principle}

As it was already pointed out in the Introduction, the universal AD
can be extracted directly from the QCD results without finding the
scalar particle contribution. This possibility is based on the deep
relation between the DGLAP and BFKL dynamics in the ${\mathcal N}=4$ SYM~
\cite{KL00,KL}.

To begin with, the eigenvalues of the BFKL kernel turn out to be
analytic functions of the conformal spin $\left| n\right| $ at least in two first orders of
perturbation theory \cite{KL}. Further, in the framework of the ${\overline{\mathrm{DR}}}$-scheme~\cite{DRED}
one can obtain from the BFKL equation (see~\cite{KL00}), that there is no
mixing among the special functions of different transcendentality levels $i$
\footnote{
Note that similar arguments were used also in~\cite{FleKoVe} to obtain
analytic results for contributions of some complicated massive Feynman
diagrams without direct calculations.},
i.e. all special functions at the NLO correction contain only sums of the
terms $\sim 1/\gamma^{i}~(i=3)$. More precisely, if we introduce the
transcendentality level $i$ for the eigenvalues $\omega(\gamma)$ of integral kernels of the BFKL
equations
in an accordance with the complexity of the terms in the
corresponding sums
(here $\Psi$ is Rimannian $\Psi$-function)
\begin{eqnarray}
\Psi \sim 1/\gamma ,~~~\Psi ^{\prime }\sim
\zeta
(2)\sim 1/\gamma ^{2},~~~\Psi ^{\prime \prime }\sim
\zeta (3)\sim 1/\gamma ^{3},
\label{trans}
\end{eqnarray}
then for the BFKL kernel in
LO and in NLO the
corresponding levels are $i=1$ and $i=3$, respectively.

Because in ${\mathcal N}=4$ SYM there is a relation between the BFKL and DGLAP equations
(see~\cite{KL00,KL}), the similar properties should be valid for the ADs
themselves, i.e. the basic functions $\gamma
_{uni}^{(0)}(j)$, $\gamma _{uni}^{(1)}(j)$ and $\gamma _{uni}^{(2)}(j)$ are
assumed to be of the types $\sim 1/j^{i}$ with the levels $i=1$, $i=3$ and
$i=5$, respectively. An exception could be for the terms appearing at a given
order from previous orders of the perturbation theory. Such
contributions could be generated and/or removed by an approximate finite
renormalization of the coupling constant. But these terms do not appear in
the ${\overline{\mathrm{DR}}}$-scheme.

It is known, that at the LO and NLO approximations
(with the SUSY relation for the QCD color factors $C_{F}=C_{A}=N_{c}$) the
most complicated contributions (with $i=1$ and $i=3$, respectively) are the
same for all LO and NLO ADs
in QCD~\cite{VMV}
and for the LO and NLO scalar-scalar ADs
\cite{KoLiVe}. This property allows one to find the
universal ADs
$\gamma _{uni}^{(0)}(j)$ and $\gamma
_{uni}^{(1)}(j)$ without knowing all elements of the AD
matrix~\cite{KL}, which was verified by the exact calculations in~\cite{KoLiVe}.

Using above arguments, we conclude, that at the NNLO level there is only one
possible candidate for $\gamma _{uni}^{(2)}(j)$. Namely, it is the most
complicated part of the QCD AD
matrix (with the SUSY relation for the QCD color factors
$C_{F}=C_{A}=N_{c}$).
Indeed, after the diagonalization of the AD
matrix its eigenvalues should have this most complicated part
as a common contribution because they differ each from others only by a shift of
the argument and their differences are constructed from
less complicated terms. The non-diagonal matrix elements of the AD
matrix contain also only less complicated terms (see, for example, AD
exact expressions at LO and NLO approximations
in Refs.~\cite{VMV}
for QCD
and~\cite{KoLiVe} for ${\mathcal N}=4$ SYM) and therefore they cannot generate
the most complicated contributions to the eigenvalues of AD
matrix.
Thus, the most complicated part of the NNLO QCD ADs
should coincide (up to color factors)
with the universal AD
$\gamma_{uni}^{(2)}(j)$.

\vskip -1cm
\section{Universal AD
for ${\mathcal N}=4$ SYM}

The final three-loop result
for the universal AD
$\gamma_{uni}(j)$
for ${\mathcal N}=4$ SYM is~\cite{KLOV}
\begin{eqnarray}
\gamma(j)\equiv\gamma_{uni}(j) ~=~ \hat a \gamma^{(0)}_{uni}(j)+\hat a^2
\gamma^{(1)}_{uni}(j) +\hat a^3 \gamma^{(2)}_{uni}(j) + ... , \qquad \hat a=\frac{\alpha N_c}{4\pi}\,,  \label{uni1}
\end{eqnarray}
where\footnote{
Note, that in an accordance with Ref.~\cite{next}
 our normalization of $\gamma (j)$ contains
the extra factor $-1/2$ in comparison with
the standard normalization (see~\cite{KL})
and differs by sign in comparison with one from Ref.~\cite{VMV}.}
\begin{eqnarray}
\frac{1}{4} \, \gamma^{(0)}_{uni}(j+2) &=& - S_1,  \label{uni1.1} \\
\frac{1}{8} \, \gamma^{(1)}_{uni}(j+2) &=& \Bigl(S_{3} + \overline S_{-3} \Bigr) -
2\,\overline S_{-2,1} + 2\,S_1\Bigl(S_{2} + \overline S_{-2} \Bigr),  \label{uni1.2} \\
\frac{1}{32} \, \gamma^{(2)}_{uni}(j+2) &=& 2\,\overline S_{-3}\,S_2 -S_5 -
2\,\overline S_{-2}\,S_3 - 3\,\overline S_{-5}  +24\,\overline S_{-2,1,1,1}\nonumber\\
&&\hspace{-1.5cm}+ 6\biggl(\overline S_{-4,1} + \overline S_{-3,2} + \overline S_{-2,3}\biggr)
- 12\biggl(\overline S_{-3,1,1} + \overline S_{-2,1,2} + \overline S_{-2,2,1}\biggr)\nonumber \\
&& \hspace{-1.5cm}  -
\biggl(S_2 + 2\,S_1^2\biggr) \biggl( 3 \,\overline S_{-3} + S_3 - 2\, \overline S_{-2,1}\biggr)
- S_1\biggl(8\,\overline S_{-4} + \overline S_{-2}^2\nonumber \\
&& \hspace{-1.5cm}  + 4\,S_2\,\overline S_{-2} +
2\,S_2^2 + 3\,S_4 - 12\, \overline S_{-3,1} - 10\, \overline S_{-2,2} + 16\, \overline S_{-2,1,1}\biggr)
\label{uni1.5}
\end{eqnarray}
and $S_{a} \equiv S_{a}(j),\ S_{a,b} \equiv S_{a,b}(j),\ S_{a,b,c} \equiv
S_{a,b,c}(j)$ are harmonic sums
\begin{eqnarray}
&&\hspace*{-1cm} S_{a}(j)\ =\ \sum^j_{m=1} \frac{1}{m^a},
\ \ S_{a,b,c,\cdots}(j)~=~ \sum^j_{m=1}
\frac{1}{m^a}\, S_{b,c,\cdots}(m),  \label{ha1} \\
&&\hspace*{-1cm} S_{-a}(j)~=~ \sum^j_{m=1} \frac{(-1)^m}{m^a},~~
S_{-a,b,c,\cdots}(j)~=~ \sum^j_{m=1} \frac{(-1)^m}{m^a}\,
S_{b,c,\cdots}(m),  \nonumber \\
&&\hspace*{-1cm} \overline S_{-a,b,c,\cdots}(j) ~=~ (-1)^j \, S_{-a,b,c,...}(j)
+ S_{-a,b,c,\cdots}(\infty) \, \Bigl( 1-(-1)^j \Bigr).  \label{ha3}
\end{eqnarray}

The expression~(\ref{ha3}) is defined for all integer values of arguments
but can be easily analytically continued to real and complex $j$
by the method of Refs. ~\cite{KK,KL}

The obtained results are very important for the verification of the
various assumptions
(see \cite{Staudacher:2004tk} and references therein)
coming from the investigations
of the properties of a conformal operators in the context of AdS/CFT correspondence~\cite{AdS-CFT}.

\section{Conclusion}
%
In this short review we presented the AD
$\gamma _{uni}(j)$ for
the ${\mathcal N}=4$ supersymmetric gauge theory up to the NNLO
approximation.
At the first three orders, the univesal AD
have been extracted from the corresponding QCD calculations.
The results for four- and fifth-loops have been obtained in
\cite{KLRSV,BaJaLu,LuReVe} from
the long-range asymptotic Bethe equations
together with some additional terms, so-called {\it wrapping
corrections}, coming in agreement with Luscher
approach.\footnote{The three- and four-loop results for the universal AD
have been reproduced
also in \cite{KoReZi}  by solution of
so-called Baxter equation, which can be obtained from
the long-range asymptotic Bethe equations.}
All the results have been obtained with using of the
{\it transcendentality principle}.\\

Author
thanks the Organizing Committee of the International
 Bogolyubov Conference
``Problems of Theoretical and Mathematical Physics"
for invitation.

\end{document}